\documentclass[11pt]{article}
\usepackage{moriond,epsfig}

\def\Journal#1#2#3#4{{#1} {\bf #2}, #3 (#4)}

\def\NPB{{\em Nucl. Phys.} B}

\def\PRL{\em Phys. Rev. Lett.}
\def\PRD{{\em Phys. Rev.} D}
\def\ZPC{{\em Z. Phys.} C}
\def\EPJC{{\em Eur. Phys. J.} C}
\def\PR{{\em Phys. Rept.}}
\def\ARNPS{{\em Ann. Rev. Nucl. Part. Sci.}}

\def\be{\begin{equation}}
\def\ee{\end{equation}}
\def\bea{\begin{eqnarray}}
\def\eea{\end{eqnarray}}

\def\lsim{\raise0.3ex\hbox{$<$\kern-0.75em\raise-1.1ex\hbox{$\sim$}}}
\begin{document}
\vspace*{4cm}

\title{NUCLEAR STRUCTURE FUNCTIONS AT SMALL $x$\\ IN MULTIPLE SCATTERING
APPROACHES}

\author{ N. ARMESTO }

\address{Theory Division, CERN\\
CH-1211 Gen\`eve 23, Switzerland}

\maketitle\abstracts{
A simple model for nuclear structure functions in the region of
small $x$ and small and
moderate $Q^2$,
is presented. It is a
parameter-free extension, in the Glauber-Gribov approach to nuclear collisions,
of a saturation model for the nucleon.
A reasonable agreement with experimental data
on ratios of nuclear structure functions
is obtained.
The unintegrated gluon
distribution and the behavior of the saturation scale
which result from this model are discussed.
}

\section{Motivation}
Nuclear structure functions are key tools to study the behavior of partons
inside nuclei. It is well known that the ratio $R(A/B)=BF_{2A}/(AF_{2B})$ of
structure functions per nucleon in different nuclei shows a complex
behavior \cite{arneodo}
depending on the region of $x$. We will be interested in the small $x$ region,
$x\lsim 0.01$ corresponding to high $\gamma^*$-nucleon energies, where
$R(A/B)<1$ (shadowing region). In this
region and in the unpolarized case, the structure function $F_2$ can be related
with the transversely or longitudinally polarized
virtual photon-target cross section,
\begin{equation}
F_{2A}(x,Q^2)=\frac{Q^2}{4\pi^2 \alpha_{\rm em}}(\sigma^A_T+\sigma^A_L).
\label{eq1}
\end{equation}
As the nucleon structure function increases very strongly with decreasing $x$,
$F_2 \propto x^{-\Delta}$, $\Delta \sim 0.2\div 0.3$, a large number of partons
is expected and non-linear effects may start to play a role.

Different approaches to this problem have been essayed (see \cite{noso} for a
brief review and \cite{cargese} for a description of the
different approaches).
On the one hand, parton densities have been parameterized at some
small $Q_0^2\gg \Lambda_{\rm QCD}^2$ and then evolved to larger $Q^2$ using
evolution equations. On the other hand, there are
models which try to describe the form
of the initial condition itself (see
e.g. \cite{meu,ja}). For this description of the small
$Q^2$ region, two alternative but equivalent pictures exist: In the rest frame
of the nucleus the virtual photon develops a hadronic $q\bar q$ fluctuation
with a coherence length which at small $x$ is larger than the nuclear size, so
multiple scattering becomes important. In
a frame in which the hadron is moving rapidly, the $q\bar q$ fluctuation sees
simultaneously the overlapping parton clouds of different nucleons in the
nucleus, and non-linear effects are expected to appear. In all these models
saturation, meaning either an upper bound in parton densities and fields or
the blackness of the scattering matrix, appears for $Q^2<Q_s^2$, with $Q_s^2$
the saturation scale.
All these studies have
great influence on the description of multiparticle production in nuclear
collisions, particularly in view of the recent RHIC data, see
\cite{cargese,ja,qm02}.

In this contribution we will describe a very simple model \cite{meu}
for nuclear structure
functions, based on the
multiple scattering picture in the dipole model \cite{dipmod}.
It is a parameter-free extension of a model for the nucleon to the
nuclear case using the Glauber-Gribov approach. The model will be presented in
the next Section, and the saturation scale it implies will be analyzed in
Section 3. Finally, some conclusions will be drawn in Section 4.

\section{Description of the model}
\begin{figure}[h]
\begin{center}
\psfig{figure=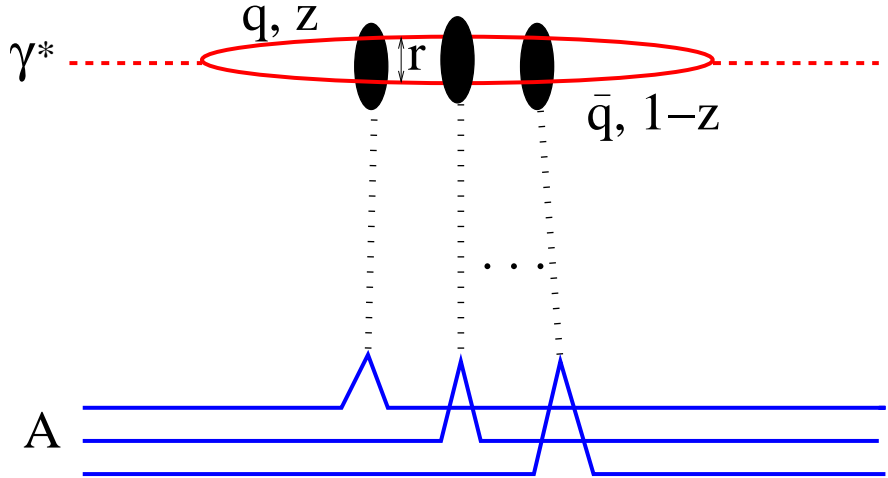,height=2.5in}
\end{center}
\caption{Interaction of the dipole fluctuation of the virtual photon with a
nuclear target.
\label{fig:dipolo}}
\end{figure}

We are going to work in the dipole model \cite{dipmod}
valid for small $x$, in which a
$q \bar q$ fluctuation of the virtual photon develops with a coherence length
$l_c\propto (xm_N)^{-1}>R_A$ and interacts hadronically with the target. In this
model (see Fig. \ref{fig:dipolo})
\begin{equation}
\sigma^A_{T,L}(x)=\int d^2r\int_0^1 dz \ |\psi_{T,L}(r,z,Q^2)|^2
\sigma_{dA}(x,r),
\label{eq2}
\end{equation}
with $\psi_{T,L}(r,z,Q^2)$ the $q\bar q$-component of the $\gamma^*$ wave
function \cite{dipmod}. For the dipole-proton cross section $\sigma_{dp}(x,r)$
we will use the model of \cite{gbw},
\begin{equation}
\sigma_{dp}(x,r)=\sigma_0 \left[1-\exp{\left(-\frac{Q_s^2(x) r^2}{4}\right)}
\right],
\label{eq3}
\end{equation}
with $Q_s^2(x)=Q_0^2\left(\frac{x_0}{\tilde{x}}\right)^\lambda$,
$\tilde{x}=x\left(
1+\frac{4m_f^2}{Q^2}\right)$, $Q_0^2=1$ GeV$^2$, $m_{u,d,s}=0.14$ GeV,
$m_c=1.5$ GeV, $\sigma_0=23.03$ (29.12) mb, $\lambda=0.288$ (0.277) and
$x_0=3.04\cdot 10^{-4}$ ($0.41\cdot 10^{-4}$) for the 3(4)-flavor
version of the model.

The extension to the nuclear case will be done in a parameter-free way using
the Glauber-Gribov approach:
\begin{equation}
\sigma_{dA}(x,r)=\int d^2b \ 2\left[1-\exp{\left(-\frac{1}{2}AT_A(b)
\sigma_{dp}(x,r)
\right)}\right].
\label{eq4}
\end{equation}
The region of validity of the model, Eqs. (\ref{eq1})-(\ref{eq4}),
is related with that of the model for
the proton \cite{gbw} and with the absence of non-linear effects:
$0.02>x> 10^{-5}\div 10^{-6}$, $Q^2<20$ GeV$^2$.
In \cite{meu} results of this model are presented. The comparison with
experimental data for different ratios $R(A/B)$ is reasonable, although at
$x\sim 0.01$ the model tends to overestimate shadowing. The nuclear effects
on the longitudinal-to-transverse cross section ratios are moderate, in any case
smaller
than 15 \% in the region of applicability of the model.

\section{Unintegrated gluon and $Q_s^2$}
\begin{figure}[ht]
\begin{center}
\psfig{figure=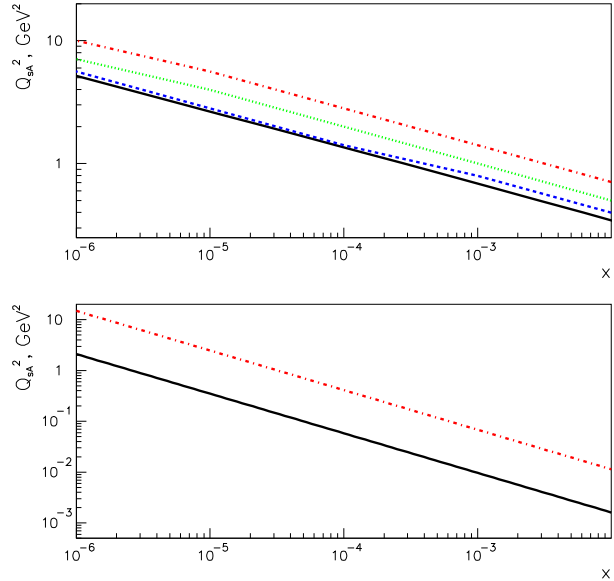,height=3.5in}
\end{center}
\caption{Upper plot: saturation momentum in the model for proton (solid line),
and for Pb
in three cases: central ($b=0$, dashed-dotted line),
peripheral ($b=7$ fm, dashed line), and integrated over
$b$ (dotted line). Lower plot: saturation momentum in the
numerical solution of the non-linear equation,
for
$A=1$ (solid line) and $A=208$ (dotted line).
Notice the difference in vertical scales between the plots.
\label{fig:fig9}}
\end{figure}

The dipole-target cross section can be related, in leading-order
$k_T$-factorization, with the unintegrated gluon distribution
\cite{smallx} $\varphi_A(x,k,b)$:
\begin{equation}
\varphi_A(x,k,b)=-\frac{N_c}{4\pi^2\alpha_s}k^2 \int \frac{d^2r}{2\pi}
\ \exp{(ik\cdot r)}
\sigma_{dA}(x,r,b),
\label{eq5}
\end{equation}
whose integral over $d^2k$ is related to the usual gluon density.
In our model, the unintegrated gluon depends on the transverse momentum $k$
just through the combination $k^2/Q_s^2$, it goes to 0 both for $k\to 0$ and
$k\to \infty$, and presents a maximum whose position in $k$ can be identified
with
the saturation scale $Q_s$. This quantity is of uttermost importance for
the applicability of models which contain saturation \cite{cargese}, as most
gluons in the wave function of the hadron get transverse momenta $k\sim Q_s$.
Thus the size of $Q_s^2$ with respect to $\Lambda_{\rm QCD}^2$ indicates
the applicability of perturbative methods in the saturation picture. In our
model the saturation scale
(for quarks, for gluons a color factor 9/4 is naively expected)
in nuclei $Q_{sA}^2$ can be related
to the saturation scale in proton $Q_s^2$. Estimations give
\begin{equation}
Q_{sA}^2\simeq
\left[4\ln{\left(\frac{2AT_A(b)\sigma_0}{2AT_A(b)\sigma_0-1}\right)}
\right]^{-1} Q_s^2\propto A^{1/3} \ \ {\rm for} \ \ A\to \infty.
\label{eq6}
\end{equation}
Numerical results are shown in Fig. \ref{fig:fig9}, compared with the results
for the saturation scale in non-linear equations \cite{am}.

Some comments are in order. First, the results of the non-linear
equations \cite{am} give a similar asymptotic $A$-dependence but a much steeper
$x$-dependence (characterized by an exponent $\lambda\simeq 0.78$ to be compared
with $\simeq 0.28$ in the model of \cite{gbw} and ours). Second, in our model
the saturation scale $Q_{sA}^2$ for central ($b=0$) Pb is 2 times the
saturation scale $Q_s^2$ for proton, notably smaller than the factor $208^{1/3}
\simeq 6$ naively expected; this result is in agreement with the conclusions
in \cite{scal}. Finally, the structure function $F_2$ in
our model shows in nuclei explicit geometrical scaling in
the variable $\tau=Q^2/Q_s^2$, at the same level as the model in \cite{gbw}
for the proton. Indeed, such scaling has been seen in proton experimental data
\cite{scalp} and searched for in nuclear data \cite{scal}; it has been
interpreted as a possible hint on the existence of a saturating, high-density
regime in small $x$, small $Q^2$ data on scattering of leptons
on nucleons or nuclei.

\section{Conclusions}
A simple model for nuclear structure functions, in the form of a parameter-free
extension of a saturating model for the proton to the nuclear case in the
Glauber-Gribov approach, has been presented. The model shows a reasonable
agreement with experimental data. It has been used to estimate the saturation
scale in nuclei, relevant for the application of saturation ideas to
multiparticle production in ultrarelativistic heavy ion collisions
\cite{cargese,qm02}. From the results of the model the saturation scale
$Q_{sA}^2$ results $<2$ GeV$^2$ for $x$-values $\sim 0.01$ relevant
for RHIC.

\section*{Acknowledgments}
The author thanks the organizers for such a nice and pleasant meeting.

\end{document}